# A novel portfolio construction strategy based on the core-periphery profile of stocks

**Imran Ansari[1*], Charu Sharma[1], Akshay Agrawal[2], Niteesh Sahni[1]**

[1]Department of Mathematics, Shiv Nadar Institution of Eminence Deemed to be University, Gautam Buddha Nagar, Uttar Pradesh-201314, India

[2]Mathematical Sciences Foundation, Greater Kailash, New Delhi-110048, India

[*] Corresponding author

Email: ia717@snu.edu.in

## Abstract

This paper highlights the significance of mesoscale structures, particularly the core-periphery structure, in financial networks for portfolio optimization. We build portfolios of stocks belonging to the periphery part of the Planar maximally filtered subgraphs of the underlying network of stocks created from Pearson correlations between pairs of stocks and compare its performance with some well-known strategies of Pozzi et. al. hinging around the local indices of centrality in terms of the Sharpe ratio, returns and standard deviation. Our findings reveal that these portfolios consistently outperform traditional strategies and further the core-periphery profile obtained is statistically significant across time periods. These empirical findings substantiate the efficacy of using the core-periphery profile of the stock market network for both inter-day and intraday trading and provide valuable insights for investors seeking better returns.

**Keywords:** Core-periphery structure, Portfolio optimization, Markowitz model, Centrality measures.

## Introduction

Portfolio optimization has been a central theme in a huge body of literature in mathematical finance over the past decades. It involves the selection of an optimal portfolio of stocks that maximizes returns while minimizing risk[1–3]. Portfolio optimization is important to high-frequency traders as well as long-term investors[4,5]. One of the first portfolio optimization model was introduced by Markowitz [6] which is essentially a quadratic optimization problem subject to a linear constraint. More precisely, we maximize the Sharpe ratio under the condition that all weights are positive and lie between 0 and 1 (in fact their sum is one)[7]. These weights represent the proportion allocated to each stock. The use of centrality and peripherality measures have emerged as effective tools to evaluate the importance of assets in a network and to construct optimized portfolios[4,5,8,9]. Many researchers have focused on the analysis of networks of stocks amid financial crises[5,10–12]. Extensive research is available on the application of centrality and peripherality measures to portfolio optimization[4,5,13,14]. It is worth pointing out that strategies other than centrality and peripherality like clustering techniques have also been explored in the context of portfolio construction[15–19].

We specifically point out the seminal work of Pozzi et al.[4] in which a measure for detecting sparsely connected vertices is constructed. This measure called the hybrid measure is based on degree centrality, betweenness centrality, eccentricity, closeness, and eigenvector centrality values. Vertices with high hybrid score are treated as peripheral, whereas vertices with low values of the hybrid score are treated as central. The underlying network used in reference[4] is the planar maximally filtered subgraph (PMFG) of the pairwise correlation-based network. The use of minimum spanning tree (MST) or PMFG subgraph of the correlation-based network has been advocated by many researchers in the recent past to formulate their investment strategies[5,13,20,21].

In Ref.[4] the authors construct portfolios with the following compositions and compare them by assigning equal weights to each constituent stock and Markowitz weights respectively to each constituent stock respectively:

- P1 comprises of $m$ most peripheral stocks. This is done by sorting the vertices in descending order of hybrid scores and then picking the stocks corresponding to the first $m$ scores.
- P2 comprises of $m$ most central stocks. This is done by sorting the vertices in ascending order of hybrid scores and then picking the stocks corresponding to the first $m$ scores.
- P3 comprises of $m$ randomly chosen stocks assuming that each stock has an equal probability of being chosen.
- P4 is the market portfolio and comprises of all the stocks listed in the market.

The striking observations made in reference[4] are that the Sharpe ratio of P1 dominate the Sharpe ratios of all other portfolios P2, P3, and P4 when equal weights are considered for each stock. But when weights are computed by solving Markowitz equations for each portfolio, then P4 outperforms all other portfolios P1, P2, and P3. Thus, the portfolio of peripheral stocks in general does not outperform the market. However, P1 always dominates P2 in Sharpe ratio. There have been many popular formulations of measures of ‘peripheralness’ based on Betweenness centrality and Eigenvector centrality respectively. In reference[4], portfolios of $m$ stocks are constructed by considering these measures in place of the hybrid measure and it is observed that P1 outperforms each of such portfolios in the context of uniform weights as well as the Markowitz weights. It is noted that peripheral stocks identified on the basis of the hybrid measure led to superior portfolios from an investment point of view as compared to the ones in which the peripheral stocks are identified on the basis of eigenvector centrality and betweenness centrality alone. The noteworthy observation in our analysis is that our proposed core-periphery-based strategy consistently outperforms P1 and P4.

There have been numerous applications of core-periphery structures across myriad domains including social networks[22–31], protein-protein interaction networks[26,28], neural networks[26,32] ,academic journals[33], and transportation networks[23,26,27,34]. The concept of core and periphery in networks was first formalized by Borgatti and Everett[22] in the context of weighted and undirected networks. Ever since it has attracted a considerable amount of interest resulting in the formulation of other viewpoints and creation of statistical and non-statistical algorithms to compute the Core and Periphery in networks[22,23,25–30,35–37].

In the present paper we shall restrict ourselves to the case of undirected graphs only. The original definition of core and periphery of a network formulated by Borgatti et. al.[22] is as follows (it captures the intuitive idea that core vertices are densely connected, there are linkages between core and periphery vertices, and there are no linkages between periphery vertices): Let $G$ be an undirected and unweighted network with $N$ vertices and $M$ edges with neither self-loop nor multiple edges. Let $A = (a_{ij})$ be the adjacency matrix for $G$ where $a_{ij} = 1$ if vertices $i$ and $j$ are adjacent, and $a_{ij} = 0$ otherwise. A network $G$ is classified as a core-periphery network if there exists a set of core vertices $K \subset N$ and periphery vertices $P \subset N \setminus K$, such that:

**i.** $\forall$ i, j $\in$ K: $a_{ij} = 1$.

**ii.** $\forall\, i, j \in P$: $a_{ij} = 0$.

**iii.** $\forall\, i \in K\ \exists\, j \in P$ with $a_{ij} = 1$, and $\forall\, j \in P\ \exists\, i \in K$ with $a_{ij} = 1$.

In Ref.[22], Borgatti et al. assigns a label $c_i$ to a vertex $i$ according to the rule: $c_i = 1$ if vertex $i$ belongs to $K$ and $c_i = 0$ if the vertex $i$ belongs to $P$; and computes the sets $K$ and $P$ by maximizing the function $\rho(c_1, \dots, c_N) = \sum_{i=1}^{N} \sum_{j=1}^{N} a_{ij} c_{ij}$ over all $c_1, \dots, c_N$. The numbers $c_{ij}$ are set to 1 if either $i$ or $j$ belong to $K$, and set to 0 if both belong to $P$. Further, in Ref.[22], the idea of assigning "coreness" value to each vertex is formulated by introducing a quantity $c_i \in [0,1]$ corresponding to each vertex $i$, and computing these by maximizing $\rho(c_1, \dots, c_N) = \sum_{i=1}^{N} \sum_{j=1}^{N} a_{ij} c_i c_j$. This way we have a method which works for networks that do not have an ideal core-periphery structure. Motivated by this approach Rombach[37] proposes a robust method capable of identifying multiple cores. In Ref.[37], quality of a core is defined in terms of a transition function and maximized to obtain the core scores for each vertex. We provide precise details in Methods section.

A literature survey shows that little work has gone into analyzing core-periphery structure in financial networks. The notable works are in the Refs.[27, 38, 39]. In Ref.[27], a combination of ETFs and their constituent stocks are studied. A core-periphery analysis of a correlation-based network using the Rombach model[37] revealed that ETFs appear in the core. No particular analysis has been reported for the periphery vertices though. In the present paper we have observed that the periphery vertices decided on the basis of the core scores given by the Rombach model was unable to give us any clear advantage from an investment point of view. So, we explored other methods of assigning core values and experimented with the method of Rossa et. al.[26]. This is a statistical procedure in which persistence probabilities of sets of vertices are computed by modelling a random walker transitioning from vertex $i$ to $j$ as a Markov chain. We provide the requisite details in Methods section.

The problem of portfolio optimization concerns selecting a combination of stocks with an objective to maximize returns or minimize risk. This task is challenging particularly in the context of the Indian stock market given its humungous size and complexity. A network-based strategy of constructing portfolios have gained popularity in the recent past[9,13,27]. In this paper, we propose a core-periphery based strategy to select optimal portfolios and compare their performance against portfolios obtained by the strategy of Pozzi et al.[4] and 'market portfolio', which consists of all stocks listed in the market. Our strategy is useful to inter-day as well as intra-day traders and the results presented in this paper are obtained on both daily and high-frequency data (recorded at intervals of 30 seconds) adding to a scarce knowledge in this context[5,10]. Our starting point is the PMFG sub-graph extracted from a pairwise correlation-based network with correlation coefficient as the weight assigned to the edge between two stocks. This is much simpler to the starting point of Pozzi's strategy in which the weights assigned to edges are the exponentially weighted Pearson's correlation coefficients[4,40].

In this paper, the results and discussion summarizes the comparison between popular portfolio construction strategies with our strategies on the basis of relevant statistical tests in addition to establishing the significance and stability of the core-periphery profile. In the remaining section we present the conclusion followed by the mathematical description and validity of the methods in our context.

## Results & Discussion

This section discusses the comparative analysis of portfolio construction, where we examine the performance of our proposed strategy in contrast to Pozzi's approach. We also analyze and compare our proposed core-periphery based strategies described in Methods section. We evaluate the performance of portfolio comprised of sizes $m = 5, 10, 20$ and $30$ stocks. We construct and compare Types-1, 2, 3, 4, 5 and 6 by assigning portfolio weights with uniform as well as Markowitz method:

Type-1 (Portfolio of periphery stocks from the core-periphery strategy-1): We apply the Rossa algorithm to each PMFG network based on correlation method. This gives a core-periphery profile: $\phi_1 \leq \phi_2 \leq \cdots \leq \phi_m \leq \cdots \leq \phi_N$, where $N$ is number of stocks. We list the pure periphery stocks for which $\alpha_k = 0$ in the decreasing order of the Sharpe ratio. Next, we construct portfolio comprising $m$ most periphery stocks corresponding to $\phi_1, \phi_2, \ldots, \phi_m$. This construction is motivated by the work in Refs.[4, 5, 8, 9, 13].

Type-2 (Portfolio of periphery stocks from the core-periphery strategy-2): We apply the Rombach algorithm to each PMFG network based on correlation method. This gives the *coreness* of each vertex and then arrange them in ascending order of coreness: $x_1 \leq x_2 \leq \cdots \leq x_m \leq \cdots \leq x_N$. Subsequently, we list the pure periphery stocks for which $x_k = 0$ in the decreasing order of the Sharpe ratio. Next, we construct portfolio comprising $m$ most periphery stocks corresponding to $x_1, x_2, \ldots, x_m$. This portfolio is constructed in a similar fashion as in Refs.[4, 5, 8, 9, 13].

Type-3 (Portfolio of peripheral stocks from the Hybrid measure of Pozzi et al.): We compute the hybrid measure for each vertex of the PMFG networks based on exponential weighted correlation method, as elaborated upon in the preliminaries section. We arrange them in descending order: $x_1 \geq x_2 \geq \cdots \geq x_N$. According to Ref.[4], peripheral vertices are the ones with higher values of hybrid measure. So, the stocks corresponding to $x_1, x_2, \ldots, x_m$ are the $m$ most peripheral stocks.

Type-4 (Portfolio of core stocks from the core-periphery strategy-1): We apply the Rossa algorithm to each PMFG network based on correlation method and arrange the obtained *core-periphery profile* in descending order. We then pick the stocks corresponding to the first $m$ scores. Thus, we are considering portfolios of $m$ most core stocks. The authors in Refs.[4, 8, 9, 13] have also considered portfolios constructed out of core stocks.

Type-5 (Portfolio of core stocks from the core-periphery strategy-2): We apply the Rombach algorithm to each PMFG network based on correlation method and arrange the obtained *coreness* in descending order. Next, we construct portfolio comprising $m$ stocks corresponding to the first $m$ scores. Thus, we are considering portfolios of $m$ most core stocks.

Type-6 (Market portfolio): We constructed 'market portfolio' comprising all $N$ listed stocks. This type of portfolio has been used as a benchmark in many works such as Refs.[4, 5].

For inter-day or intraday traders, investing in a large portfolio of stocks can be challenging due to the high volume of transactions and the need for rapid decision-making. Therefore, our objective was to develop an investment strategy that enables investors to achieve competitive returns with a small portfolio of stocks compared to a fully diversified portfolio containing a large number of stocks. Therefore, we evaluate the performance of portfolios constructed with varying sizes, specifically m=5, 10, 20, and 30 comprising the most periphery and core stocks using our proposed core-periphery-based strategies. In the subsequent section, we will elucidate the portfolio performance of our proposed strategies utilizing both daily and high-frequency stock market datasets.

**A comparative study on daily time series stock market data:** We utilized daily data from 351 stocks selected from the NIFTY 500 index of the National Stock Exchange (NSE) of India spanning from January 2014 to December 2021, covering a total of 1970 market days. Notably, during this period, the market trend experienced a significant downturn (Supplementary Fig. S1) in 2020 due to the coronavirus disease 2019 (COVID-19) pandemic, resulting in prolonged volatility. Our analysis involves utilizing moving windows of 125 days (equivalent to 6 months). For each day $t$ within this window, we construct the PMFG-filtered network based on the full cross-correlation matrix derived from the preceding 125 days. Subsequently, we assess the stability of the portfolio over the subsequent 125 days, encompassing holding periods ranging from 1 to 125 market days.

We first construct portfolios using the uniform method and then compare the Sharpe ratios and average returns of Types-1, 2, 3, 4, 5 and 6 in the daily time series data (NIFTY 500). Supplementary Figs. S2-S3 show that Type-1 significantly outperforms all Types-2, 3, 4, 5 and 6 in terms of both the Sharpe ratio and average returns. Notably, as the holding period increases, Type-1 consistently demonstrates increasing advantages over Types-2, 3, 4, 5 and 6. Our main results are presented by the comparison between Type-1 with Type-3 and Type-6 (investing in all stocks i.e., market performance). Furthermore, Supplementary Fig. S4 indicates that the risk (standard deviation) associated with Type-1 is lower than that of Types-4 and 5, while as portfolio size increases risk profiles of Type-1 becomes similar to Types-3 and 6. We also construct portfolios using the Markowitz method and then compare the Sharpe ratios and average returns of Types-1, 2, 3, 4, 5 and 6. Fig. 1 and Supplementary Fig. S5 show that Type-1 significantly outperforms Types-2, 3, 4 and 5 in terms of both the Sharpe ratio and average return. Significantly, as the holding period extends, Type-1 consistently demonstrates increasing advantages over Types-1, 2, 3, 4, and 5. Additionally, from Fig. 1, as the portfolio size expands, Type-1 approaches Type-6, representing the market portfolio with Markowitz weights, in terms of the Sharpe ratio. However, analysis from Supplementary Fig. S5 reveals that the average return of Type-1 is comparable to that of Type-6. Furthermore, Supplementary Fig. S6 indicates that the risk, as measured by standard deviation, associated with Type-1 is lower than that of Types-2, 3, 4, and 5. Additionally, as the portfolio size increases, the risk profiles of Type-1 become increasingly similar to those of Type-6. This disparity in risk for Type-6 may be attributed to the allocation of Markowitz weights across all stocks.

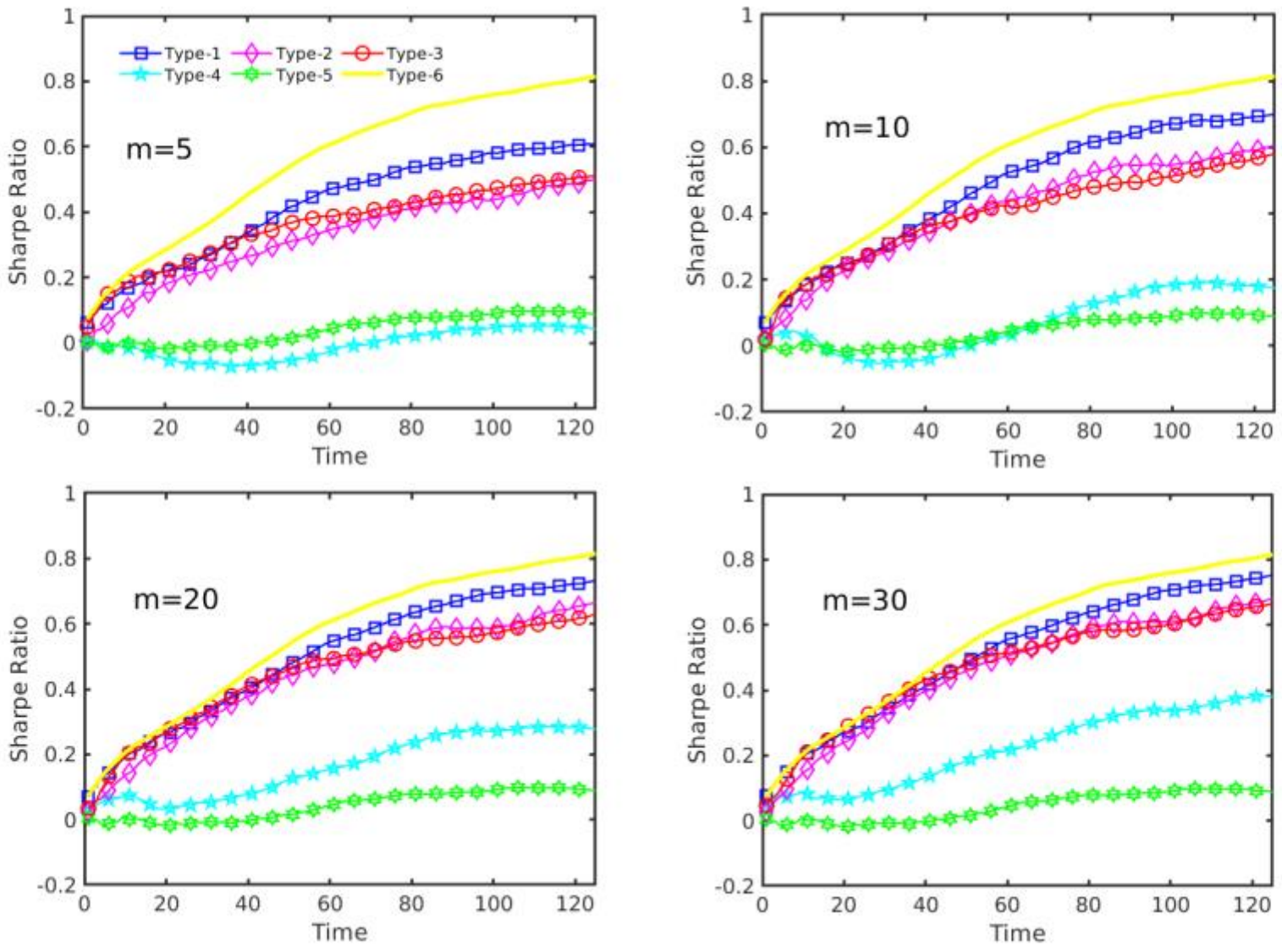

**Figure 1.** Each subplot compares the Sharpe ratio of portfolios of Types-1, 2, 3, 4, 5 and 6 of sizes 5, 10, 20 and 30 stocks respectively; for the following holding periods, including 1, 2, and so on, up to 125 days; weights assigned through the Markowitz method.

**A comparative study on high-frequency stock market data:** We conduct our analysis using 30-second tick-by-tick high-frequency data obtained from the NSE (India), covering the period from January 2014 to December 2014, comprising a total of 164,640 ticks. We specifically focused on the year 2014 due to the significant political event of the general elections in India, which resulted in a change of government from the United Progressive Alliance (UPA) to the National Democratic Alliance (NDA) after a decade. The elections, promotional rallies held in March, polling in April, and result declarations in May collectively contributed to heightened market volatility during this period, which persisted post-election (Supplementary Fig. S7). Our analysis involves utilizing moving windows of 240 ticks (equivalent to 2 hours) for the high-frequency data. At each instant t, we construct the PMFG filtered network using 75% overlapping windows of size 240 ticks. Due to the extensive size of the dataset, we employ overlapping windows to efficiently construct PMFG networks. Subsequently, we assess the stability of the portfolio over the next 2-hour window, considering holding periods ranging from 1 to 240 ticks. Initially, portfolios are construct employing the uniform method, followed by a comprehensive comparison of the Sharpe ratios and average returns across various types (Types-1, 2, 3, 4, 5, and 6) within the high-frequency time series data (CNX100). Supplementary Figs. S8-S9 illustrate that Type-1 exhibits a significant outperformance compared to Types-2, 3, 4, 5, and 6 in terms of both the Sharpe ratio and average returns. Importantly, as the holding period extends, Type-1 consistently demonstrates escalating advantages over Types-2, 3, 4, 5, and 6. Our main results are presented by the comparison between Type-1 with Types-3 and 6 (market portfolio). Moreover, Supplementary Fig. S10 illustrates that the risk, as indicated by standard deviation, associated with Type-1 is lower compared to that of Types-4 and 5. However, as the portfolio size increases, the risk profiles of Type-1 become increasingly similar to that of Type-6. We also constructed portfolio using the Markowitz method, followed by a comparative analysis of the Sharpe ratios and average returns across Types-1, 2, 3, 4, 5, and 6. Fig. 2 and Supplementary Fig. S11 show that Type-1 significantly outperforms Types-2, 3, 4 and 5 in terms of both the Sharpe ratio and average return. Significantly, as the holding period lengthens, Type-1 consistently exhibits increasing advantages over Types-1, 2, 3, 4, and 5. Additionally, according to Fig. 2, as the portfolio size expands, Type-1 converges toward Type-6, representing the market portfolio with Markowitz weights, in terms of the Sharpe ratio. However, contrasting insights from Supplementary Fig. S11 reveal that the average return of Type-1 outperforms that of Type-6. Moreover, Supplementary Fig. S12 highlights that the risk, as indicated by standard deviation, associated with Type-1 is lower than that of Types-2, 3, 4, and 5. Notably, Type-6 appears less risky, potentially attributed to the allocation of Markowitz weights across all stocks. It's worth emphasizing that the Markowitz theory serves as a classical mathematical framework aimed at maximizing the Sharpe ratio in portfolio construction.

Finally, our study reveals that portfolios with Markowitz weights exhibit a lower Sharpe ratio compared to those with uniform weights, aligning with the classical framework of the Markowitz method. While the Markowitz method theoretically applies to a large number of stocks, practical implementation can be challenging and costly, making our method more appealing for financial applications. Our results remain consistent across both intraday and intraday investor types, demonstrating the robustness of our approach. By analyzing both uniform and Markowitz-weighted portfolios, we conclude that our strategy outperforms in both portfolio types. Overall, our findings underscore the advantage of mesoscale structures in portfolio construction and highlight the significance of periphery stocks in achieving superior risk-adjusted returns.

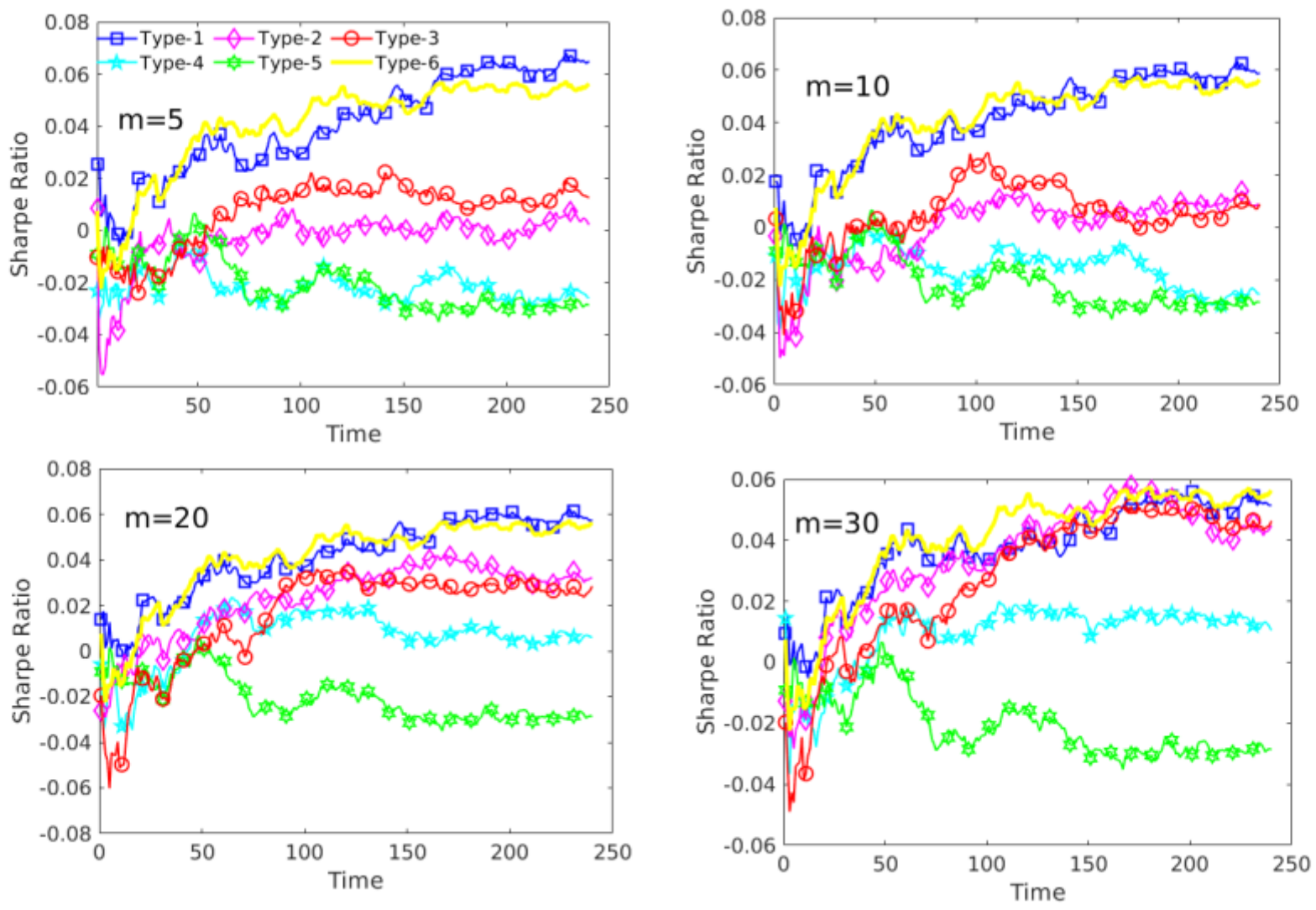

**Figure 2.** Each subplot compares the Sharpe ratio of portfolios of Types-1, 2, 3, 4, 5 and 6 of sizes 5, 10, 20 and 30 stocks respectively; for the following holding periods, including 30 seconds, 1 minute, 1.5 minutes, and so on, up to 120 minutes (total 240 ticks); weights assigned through the Markowitz method.

**Cross validating our observations:** In order to cross validate and check the effectiveness of our method, we randomly picked 500 windows (out of 1846 windows) from the daily time series data set (NIFTY 500). Then picked stocks based upon strategies of Types-1, 2 and 3. Subsequently we compared the rate of returns of the portfolios for different holding periods (for example, 50, 60, and so on, up to 125 days). We compare the Sharpe ratios, average rate of returns and standard deviations for different strategies (Types-1, 2 and 3). The process was repeated 1000 times and obtained average return, standard deviation, and Sharpe Ratio for each 1000 iteration corresponding to Types-1,2 and 3. We carried out hypothesis testing in support of our claim that the portfolio comprising of the Type-1 outperforms the others (Types-2 and 3). We calculated $\hat{p}$, the proportion of times Sharpe Ratio from our proposed strategy was more than the Sharpe Ratio of the other strategies and carried out hypothesis testing for the proportion test as follows:

Null Hypothesis ($H_0$): $p_0 = 0.7$.

Alternative Hypothesis ($H_a$): $p_0 > 0.7$.

Next, we calculated the p-value on the basis of following test statistic

$$z = \frac{\hat{p} - p_0}{\sqrt{\frac{p_0(1 - p_0)}{1000}}}$$

For small $p$-value we reject the null hypothesis and accept the alternative hypothesis.

Similar tests were carried out to check the proportion of times average return from Type-1 strategy more than Types-2 and 3, alongside evaluating whether the standard deviation of Type-1 strategy is lower than that of Types-2 and 3. The p-values for Sharpe ratio comparisons between Types-1 and 2, as well as Types-1 and 3, for various holding periods, are presented in Tables 1 and 2, respectively. One can clearly claim that at least 70% of the randomly chosen windows our proposed strategy (Type-1) performed better than the rest of the strategies (Types-2 and 3). The p-values for the comparisons of average returns and standard deviations between Types-1 and 2, as well as Types-1 and 3, across various holding periods, are provided in the Supplementary Tables S1-S4, respectively.

| Holding period ($T$) | 5 stocks | | 10 stocks | | 20 stocks | | 30 stocks | |
|---|---|---|---|---|---|---|---|---|
| | u | m | u | m | u | m | u | m |
| 50 | 0 (0.97) | 0 (0.96) | 0 (0.96) | 0 (0.83) | 0 (0.97) | 0.005 (0.73) | 0 (0.95) | 0.7 (0.68) |
| 60 | 0 (0.98) | 0 (0.98) | 0 (0.98) | 0 (0.93) | 0 (0.99) | 0 (0.9) | 0 (0.97) | 0 (0.81) |
| 70 | 0 (0.97) | 0 (0.97) | 0 (0.99) | 0 (0.95) | 0 (0.99) | 0 (0.91) | 0 (0.97) | 0 (0.84) |
| 80 | 0 (0.98) | 0 (0.98) | 0 (0.99) | 0 (0.95) | 0 (0.99) | 0 (0.91) | 0 (0.98) | 0 (0.84) |
| 90 | 0<br>(0.98) | 0<br>(0.98) | 0<br>(0.99) | 0<br>(0.95) | 0<br>(0.98) | 0<br>(0.92) | 0<br>(0.97) | 0<br>(0.89) |
| 100 | 0<br>(0.97) | 0<br>(0.99) | 0<br>(0.98) | 0<br>(0.98) | 0<br>(0.98) | 0<br>(0.97) | 0<br>(0.98) | 0<br>(0.95) |
| 110 | 0<br>(0.94) | 0<br>(0.98) | 0<br>(0.96) | 0<br>(0.97) | 0<br>(0.97) | 0<br>(0.96) | 0<br>(0.96) | 0<br>(0.92) |
| 125 | 0<br>(0.87) | 0<br>(0.95) | 0<br>(0.89) | 0<br>(0.95) | 0<br>(0.92) | 0<br>(0.89) | 0<br>(0.88) | 0<br>(0.9) |

**Table 1.** Hypothesis testing results for proportion of times **Sharpe ratio** of Type-1 are higher than Type-2 (p-values < 0.05, rejecting null hypothesis in favour of alternative). **Clearly Type-1 performs better than Type-3.** In the round brackets the proportion of the times Type-1 outperforms Type-2 in Sharpe ratio. The symbols in the first column denote the holding period, while those in the first row represent the portfolio size, where m = 5, 10, 20, and 30, respectively. The table reports results for portfolios constructed using both uniform (u) and Markowitz weights (m).

| Holding period ($T$) | 5 stocks | | 10 stocks | | 20 stocks | | 30 stocks | |
|---|---|---|---|---|---|---|---|---|
| | u | m | u | m | u | m | u | m |
| 50 | 0<br>(0.95) | 0<br>(0.82) | 0<br>(0.96) | 0<br>(0.83) | 0<br>(0.97) | 0.70<br>(0.69) | 0<br>(0.96) | 0.90<br>(0.66) |
| 60 | 0<br>(0.98) | 0<br>(0.95) | 0<br>(0.98) | 0<br>(0.92) | 0<br>(0.99) | 0<br>(0.89) | 0<br>(0.98) | 0<br>(0.86) |
| 70 | 0<br>(0.97) | 0<br>(0.97) | 0<br>(0.99) | 0<br>(0.97) | 0<br>(0.99) | 0<br>(0.93) | 0<br>(0.99) | 0<br>(0.90) |
| 80 | 0<br>(0.98) | 0<br>(0.97) | 0<br>(0.99) | 0<br>(0.98) | 0<br>(1.0) | 0<br>(0.96) | 0<br>(0.99) | 0<br>(0.91) |
| 90 | 0<br>(0.95) | 0<br>(0.96) | 0<br>(0.99) | 0<br>(0.99) | 0<br>(1.0) | 0<br>(0.98) | 0<br>(0.99) | 0<br>(0.97) |
| 100 | 0<br>(0.92) | 0<br>(0.97) | 0<br>(0.99) | 0<br>(0.99) | 0<br>(1.0) | 0<br>(0.99) | 0<br>(0.99) | 0<br>(0.98) |
| 110 | 0<br>(0.82) | 0<br>(0.97) | 0<br>(0.87) | 0<br>(0.99) | 0<br>(0.99) | 0<br>(0.99) | 0<br>(0.98) | 0<br>(0.98) |
| 125 | 7.8E-08<br>(0.77) | 0<br>(0.98) | 0<br>(0.95) | 0<br>(0.98) | 0<br>(0.99) | 0<br>(0.98) | 0<br>(0.97) | 0<br>(0.96) |

**Table 2.** Hypothesis testing results for proportion of times **Sharpe ratio** of Type-1 are higher than Type-3 (p-values < 0.05, rejecting null hypothesis in favour of alternative). **Clearly Type-1 performs better than Type-3.** In the round brackets the proportion of the times Type-1 outperforms Type-3 in Sharpe ratio. The symbols in the first column denote the holding period, while those in the first row represent the portfolio size, where m = 5, 10, 20, and 30, respectively. The table reports results for portfolios constructed using both uniform (u) and Markowitz weights (m).

**Statistical significance of core-periphery profile of stocks:** For the statistical analysis of core-periphery structure we divide our dataset of CNX100 index into two-hour windows throughout the entire year 2014. There was a total of 686 windows in the entire year. In Ref.[26], a measure of the strength of the core-periphery structure called the core-periphery centralization (cp-centralization), is provided and is given by

$$C = 1 - \frac{2}{n-2}\sum_{k=1}^{n-1} \varphi_k.$$

A high value of $C$ indicates that the obtained core-periphery profile $(\phi_1, \dots, \phi_n)$ is significant.

A statistical procedure to judge the significance of the observed value of $C$ is based on the computation of the $p$-value calculated as follows: For each network we generate 100 randomized networks which preserve the same degree distribution as the original network[45]. We calculate the cp-centralization for each randomized network and call it as $C_i^{rand}, i = 1,2,\dots,100$. Next, we choose between the following hypotheses on the basis of $p$-values calculated as $\frac{\#\{i: C_i^{rand} > C\}}{100}$:

Null Hypothesis $(H_0)$: The profile $(\phi_1, \dots, \phi_n)$ is not significant.

Alternative Hypothesis $(H_a)$: The profile $(\phi_1, \dots, \phi_n)$ is significant and not random.

For small $p$-value we reject the null hypothesis $H_0$ and accept the alternative hypothesis. 77% of the 686 network windows were found to have a significant core-periphery profile at 5% level of significance($p$-value<0.05) and at the same time 81% of the network windows were found to have a significant core-periphery profile at a 10% level of significance($p$-value<0.1). We summarize the statistics in Supplementary Table S5: the cp-centralization ($C$), the mean and standard deviation of the cp-centralization $C_i^{rand}$ for each randomized network of all 686 networks. In Fig. 3, we plot the histogram of cp-centralization values obtained for the 686 networks constructed over a 1-year time span.

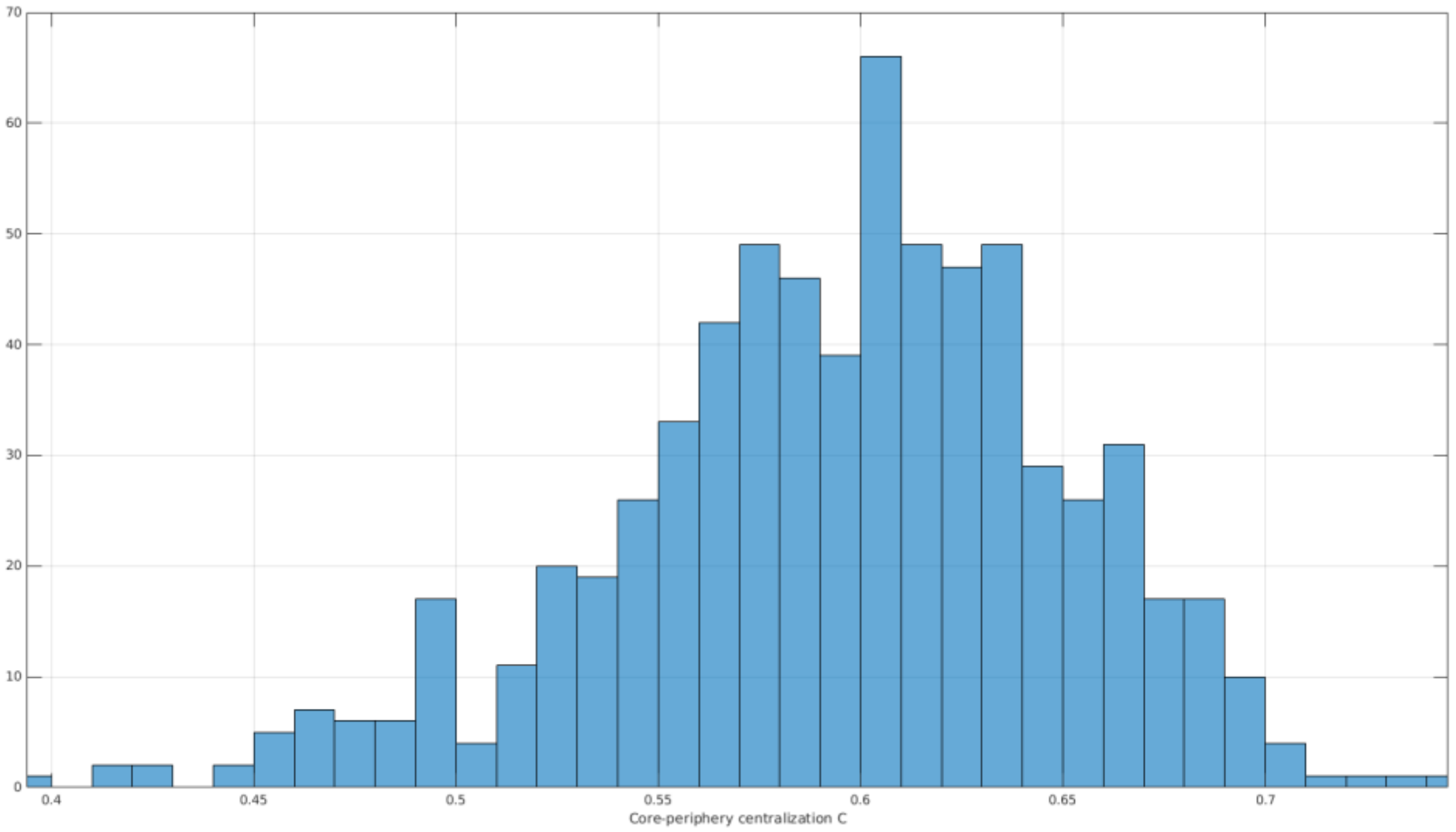

**Figure 3.** Histogram of cp-centralization $C$ for all 686 network windows.

**Dynamic occurrence of core and periphery stocks over time:** To explore the dynamic occurrence patterns of core and periphery stocks, we conducted a comprehensive analysis encompassing all 351 stocks within the Nifty 500 index. Specifically, we investigated the top 20 core and top 20 periphery stocks across 1846 windows, employing the method outlined by Rossa et al. detailed in Methods section. The accompanying figure provides a detailed depiction of stock behavior across these time windows, revealing discernible differences between core and periphery stocks. Each horizontal line in the Fig. 4 signifies the continuous selection of a particular stock within the respective time window. A striking observation is the relatively stable occurrence of the core stocks throughout the considered period. In contrast, the periphery stocks are seen to exabit a very behavior. This stability suggests that core stocks are consistently in central positions over time, whereas periphery stocks display greater variability.

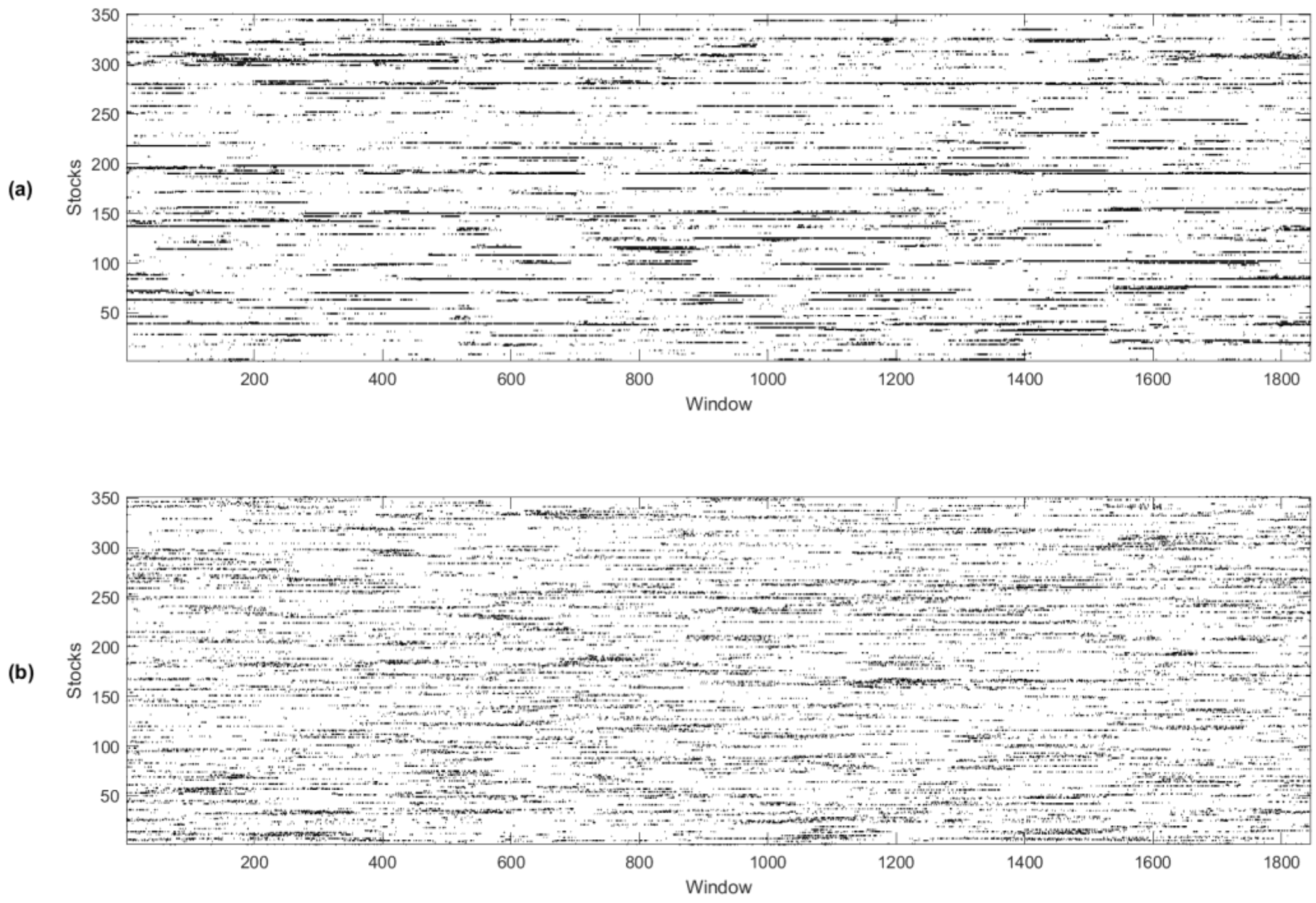

**Figure 4.** The plot shows occurrences of the top 20 core and periphery stocks across all 1846 windows of the Nifty 500 index: (a) Core Stocks: Dots represent stock selection within the top 20, with horizontal lines indicating continuous selection. (b) Periphery Stocks: Like core stocks, dots denote stock selection within the top 20, with horizontal lines representing continuous selection.

We have conducted a comprehensive analysis to ascertain the occurrence distribution, which denotes the number of days a stock is allocated to either the core or periphery group. Through this analysis, we computed the distribution of occurrence times for individual stocks within the portfolio. The results of this analysis are presented in Fig. 5. The distribution of occurrence times for core stocks reveals a notable peak at the shortest occurrence times, with a pronounced tail extending towards longer durations. Notably, the distribution of core stocks demonstrates characteristics akin to a power law distribution, as evidenced in the Fig. 5. For instance, a discernible pattern emerges where 15 stocks have consistently remained part of the core for a duration exceeding 500 days. In contrast, the distribution pattern for periphery stocks follows an exponential decay trend. This disparity in distribution behaviors between core and periphery stocks underscores the nuanced dynamics within the portfolio.

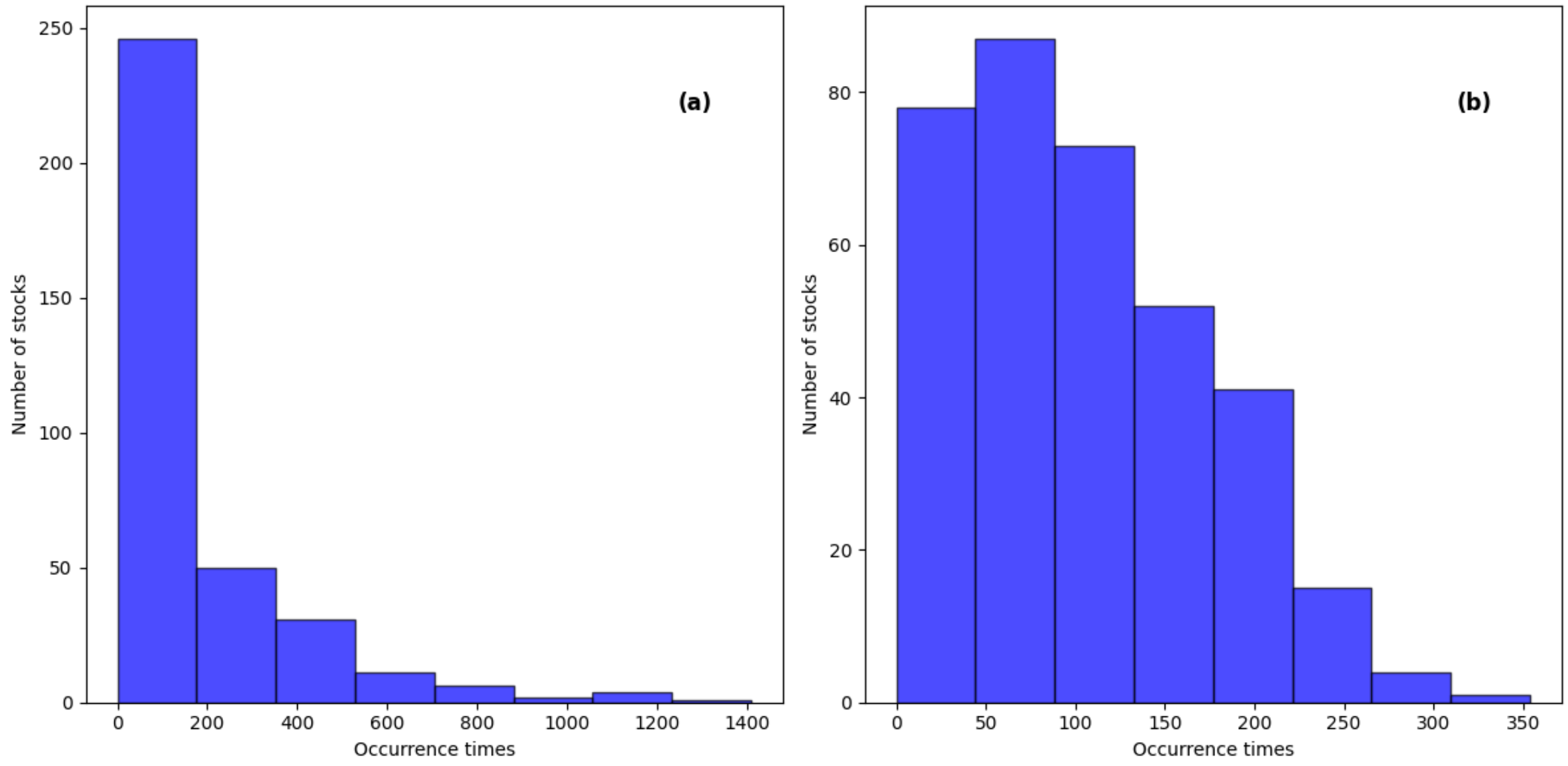

**Figure 5.** Histogram displays the occurrence times of the top 20 core and top 20 periphery stocks in all 1846 windows of the Nifty 500 index assigned to (a) core and, (b) periphery stocks.

We record the frequency of all 351 stocks of Nifty 500 occurring in the top 20 core and the top 20 peripheral stocks across all 1846 windows. We observe that core stocks are predominantly concentrated in financial services. Additionally, Metals & Mining, Chemicals, and construction materials also have stocks with high coreness values. Specifically, the most occurring core stocks across 1846 windows were L&T Finance Holdings Ltd. (Financial Services) (70.10%), Bank of India (Financial Services) (56.07%), Steel Authority of India Ltd. (Metals & Mining) (52.11%), Canara Bank (Financial Services) (48.86%). In contrast, the peripheral companies span various sectors. Specifically, the largest occurring periphery stocks across all 1846 windows were Abbott India Ltd. (Healthcare) (20.21%), Tanla Platforms Ltd. (Information Technology) (15.76%), Relaxo Footwears Ltd. (Consumer Durables) (15.28%), RattanIndia Enterprises Ltd. (Services) (15.01%), APL Apollo Tubes Ltd. (Capital Goods) (14.57%), Alkyl Amines Chemicals Ltd. (Chemicals) (14.41%). Therefore, the core stocks are stable in various time windows whereas periphery stocks have larger variations.

In terms of the industrial sectors the frequently occurring core stocks across 1846 windows belongs to Financial Services (28.43%), Metals & Mining (11.53%), Chemicals (7.77%), Construction Materials (7.10%), Automobile and Auto Components (6.45%), Capital Goods (5.74%). The largest occurring periphery populated sectors are Healthcare (12.30%), Capital Goods (11.16%), Chemicals (10.13%), Financial Services (9.76), Consumer Durables (9.04%), Fast Moving Consumer Goods (7.64%). Fig. 6 summarizes the proportion of allocation of occurrences of the core and the peripheral stocks.

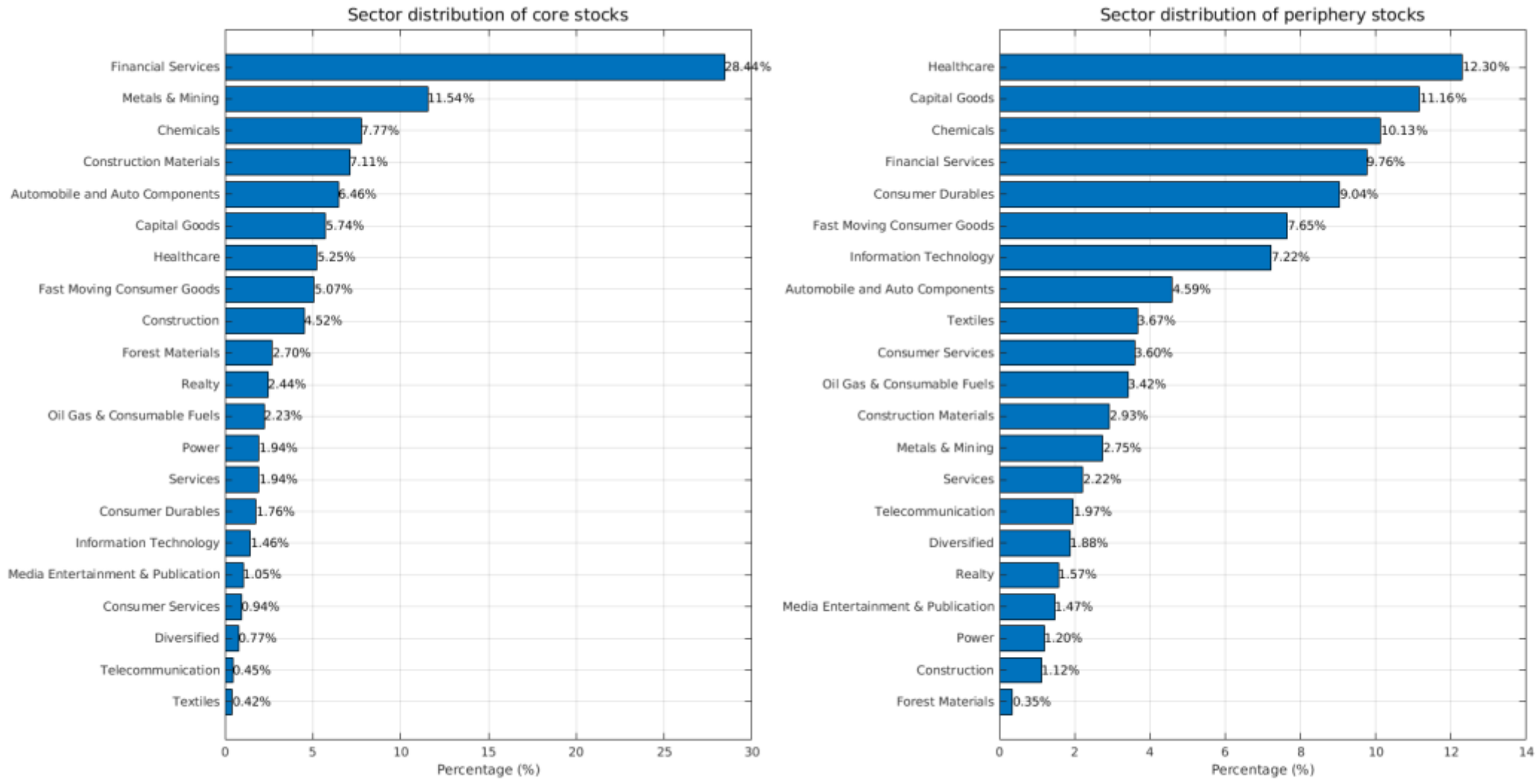

**Figure 6.** Allocation of weights to sectors based on the frequency of occurrences of a stock in the top 20 core and periphery stocks corresponding to all 1846 windows.

## Conclusion

In previous research endeavors, researchers have delved into portfolio optimization through the lens of network centralities and peripheralities, thereby addressing high-order interconnection risks. In this paper, we present an innovative investment strategy inspired by the mesoscale structures inherent in networks for portfolio optimization purposes. We work with both high frequency and daily price returns time series data sets. We build the financial complex network of stocks with the correlation method using the network filtering approach, namely PMFG. It has been shown that the core-periphery structures of the PMFG subgraphs across various time windows are statistically significant. Our study reveals that our proposed core-periphery-based strategy consistently outperforms traditional methods, resulting in higher returns and lower risk. Specifically, portfolios constructed from stocks within the periphery segment of the PMFG subgraph, exhibit superiority over portfolios constructed from the large hybrid measure (that is peripheral in the sense of Ref.[4]). It is further observed that the underlying network that is being analyzed for core and periphery parts has relatively simpler edge weights (in fact, correlation coefficients) as opposed to exponential weights in case of networks constructed in Ref.[4]. The periphery part arising on the basis of persistence probabilities[26] gives far superior results in comparison with the periphery part arising on the basis of core scores[37]. Furthermore, we show that investing in the periphery stocks perform much better than the core stocks and competing with the market portfolio. Our results are consistent with both uniform as well as Markowitz method. Our method has been tested on high frequency as well as daily stocks market data and so it is of value to both inter-day as well as intra-day traders.

# Methods

## Preliminaries

**Planar Maximally Filtered Graph:** Planer maximally filtered graph (PMFG) is a network filtering approach introduced in[41]. It works by retaining the key representative links and has shown to be an applicable tool for filtering out the most relevant information from the networks, especially in correlation-based networks[4,8,43].

Let $G$ represents a graph with a set $V(G)$ of $N$ vertices. A graph G is planar if G can be embedded in the plane in such a way that two edges intersect only at their endpoints. According to Kuratowski's theorem, a finite graph is planar if and only if it does not have a subgraph that is homeomorphic to $K_5$ (complete graph on 5 vertices) or $K_{3,3}$ (complete 3-3 bipartite graph). A planar graph is called maximal planar if adding any additional edges would make it non-planar. The construction of the PMFG subgraph begins by sorting the elements of the upper triangular part of the adjacency matrix (of order $N$) in descending order. In the first iteration, we pick the first 6 elements from this sorted list and retain all the vertices and edges corresponding to these elements of the sorted list. Call this graph as PMFG. In the next iteration, include the vertices and edges corresponding to the $7^{th}$ element of the list and test the planarity condition (using Boyer–Myrvold planarity test[42]). If the condition is met, then PMFG is updated to this new graph. This process is repeated till we exhaust all the elements in the sorted list and in the end, we will be left with PMFG having $3(N-2)$ edges. When this process ends, we are left with a maximally filtered planar graph.

**Weighted Pearson Correlation:** Let $(y_1^k, y_2^k, \dots, y_T^k)$ be the time series of size $T$ representing log returns of prices for stocks $k$. Then the pairwise weighted Pearson's correlation is defined by

$$\rho_{ij}^w = \frac{\sigma_{ij}^w}{\sigma_i^w \sigma_j^w},$$

where $\sigma_k^w, \sigma_{ij}^w$ are the weighted sample standard deviation and covariance respectively, which are defined as:

$\sigma_{ij}^w = \sum_{t=1}^T w_t(y_t^i - \bar{y}_i^w)(y_t^j - y_{avg}^j)$, and ${\sigma_k^w}^2 = \sum_{t=1}^T w_t(y_t^k - y_{avg}^k)^2$, here $y_{avg}^k = \sum_{t=1}^T w_t\, y_t^k$ represents weighted sample mean.

In Refs.[4,40] the exponential weights $w$ are chosen according to the prescription:

With the constraint that $w_t > 0 \; and \; \sum_{t=1}^T w_t = 1$. In our case, $T = 240$ and $\theta = 240$. The initial weight $w_0$ can be computed from the constraint $\sum_{t=1}^T w_t = 1$.

$$w_t = w_0 \exp\left(\frac{t-T}{\theta}\right), \forall\, 1 \le t \le T \tag{1}$$

The rationale of choosing equation (1) in Ref.[4] is to assign greater weight to more recent observations.

**Hybrid Measure and Sharpe Ratio:** The approach adopted in Ref.[4] is to calculate the hybrid measure for each vertex of the PMFG subgraph extracted from the full exponential weighted Pearson correlation network. This hybrid measure is defined in terms of Degree centrality (DC), the Betweenness (BC), Eccentricity(E), the Closeness(C), and the Eigenvector centrality (EC) for the underlying weighted and unweighted PMFG subgraphs. The hybrid measure of the $i^{th}$ vertex is defined by

$$P(i) = \frac{C_{DC}^w(i) + C_{DC}^u(i) + C_{BC}^w(i) + C_{BC}^u(i) - 4}{4(n-1)} + \frac{C_E^w(i) + C_E^u(i) + C_C^w(i) + C_C^u(i) + C_{EC}^w(i) + C_{EC}^u(i) - 6}{6(n-1)}, \tag{2}$$

For the vertex $i$, $C_{DC}^w(i)$ ($C_{DC}^u(i)$) is the weighted (unweighted) degree centrality. $C_{BC}^w(i)$ ($C_{BC}^u(i)$) is the weighted (unweighted) betweenness centrality. $C_E^w(i)$ ($C_E^u(i)$) is the weighted (unweighted) eccentricity. $C_C^w(i)$ ($C_C^u(i)$) is the weighted (unweighted) closeness. $C_{EC}^w(i)$ ($C_{EC}^u(i)$) is the weighted (unweighted) eigenvector centrality. For definitions of these quantities, the reader can refer to Ref.[43]. In Ref.[4], for the computation of weighted degree and eigenvector centrality of vertices, the edge between vertices $i$ and $j$ are assigned the weight $1 + \rho_{ij}^w$ ; whereas, for the computation of weighted betweenness, eccentricity, and closeness centralities the edge weights chosen are $\sqrt{2(1-\rho_{ij}^{w})}$. The value of $P(i)$ is small for the central vertices and large for its peripheral vertices in the network. Supplementary Fig. S13 shows the correlation based PMFG network of 89 stocks of the index CNX100 for the entire year 2014. The color of the vertices is according to the hybrid measure given by equation (2).

Let $\hat{S}_{t,m}$ represent the price of the portfolio of the $m$ stocks at tick t. Let $w_i$ be the weight of the $i^{th}$ stock in the portfolio then for each portfolio at tick $t$ , the return of the portfolio over the next T ticks is given by,

$$R_{t+T,m} = \sum_{i=1}^{m} w_i R_{t+T,i}$$

with

$$R_{t+T,i} = ln \frac{\hat{S}_{t+T,i}}{\hat{S}_{t,i}}$$

where $\hat{S}_{t,i}$ and $\hat{S}_{t+T,i}$ be the price of the $i^{\text{th}}$ stock at tick $t$ and $t+T$ respectively. In this context, T is referred to as the holding period. We compute the average return $\bar{R}_{t+T,m}$, as well as standard deviation $s_{T,m}$ over $t$ in the full-time span. We then chose the Sharpe ratio $S_{T,m}$ as a proxy for the performance of portfolio which is defined as

$$S_{T,m} = \frac{\bar{R}_{t+T,m}}{s_{T,m}}.$$

From an investment point of view, it is desirable to invest in portfolios having large Sharpe ratios.

**Markowitz model:** The Markowitz mean-variance portfolio theory, proposed by Markowitz[6], is a fundamental framework in modern finance for optimizing portfolio allocations. In this study, we use the Markowitz model that maximizes the Sharpe ratio, which measures the risk-adjusted return of a portfolio[7]. The objective is to find the optimal weights of assets that maximize the Sharpe ratio, subject to certain constraints. Let $w$ be the weight vector representing the allocation of assets in the portfolio. The optimal portfolio allocation can be obtained by solving the following optimization problem:

$$\max_{w} \frac{R_P}{\sigma_P}$$

subject to $\sum_i w_i = 1\ and\ w_i \geq 0, \forall\ i$

where $R_P$ is the expected return of the portfolio and $\sigma_P$ is the standard deviation of the portfolio returns. The expected return of the portfolio $R_P$ and the standard deviation $\sigma_P$ can be calculated as follows: $R_P = \sum_i w_i . R_i$, $\sigma_P = \sqrt{\sum_i \sum_j w_i . w_j . \sigma_i \sigma_j . C_{ij}}$ where $R_i$ is the expected return of asset $i$, $\sigma_i$ is the standard deviation of the returns of asset $i$, and $C_{ij}$ is the cross-correlation between assets $i$ and j.

## Methodology

**Data set 1:** We collect tick by tick high frequency data of the constituent stocks of the $\mathrm{CNX100}$ index from the National Stock Exchange, India during the period January $2014$ to December $2014$. The data was filtered to obtain all stocks listed on the $\mathrm{CNX100}$ during that year. 11 stocks were omitted from the analysis due to insufficient and missing data. We finally selected 89 stocks out of 100 stocks during the entire year 2014. The exchange opens at 9 am and is open till 4 pm. In the first half hour trades tend to pick up pace, while the last half hour shows some ambiguity or incompleteness in the data. Keeping this in mind, we have used data from $9:30\ am$ to $3:30\ pm$ in our analysis. Furthermore, we divide this period into 30 seconds time intervals and call each such interval a tick. Thus, the total number of ticks considered for each working day will be $720$. For the $k^{th}$ stock, we first calculate the volume weighted average price, $\hat{S}_{t,k}$ for the tick t by

$$\hat{S}_{t,k} = \frac{\sum_i v_{i,k}^t S_{i,k}^t}{\sum_i v_{i,k}^t}$$

Here $v_{i,k}^t$ is the volume of the $k^{th}$ stock traded at an instant $i$ and $S_{i,k}^t$ is the stock price at the instant $i$ in 30-second window at time $t$. The log return of each stock k at tick $t$ is then calculated as:

$$R_{t+1,k} = ln \frac{\hat{S}_{t+1,k}}{\hat{S}_{t,k}}$$

Further, we considered the year 2014 for our analysis as it was year when general elections were held in India and a change in government was seen. Table 3 provides detailed information about sector-wise distribution of the 89 stocks in the CNX 100 index considered in our analysis.

| Industry | # of stocks |
|---|---|
| Industrial Manufacturing | 5 |
| Cement & Cement Products | 5 |
| Services | 2 |
| Automobile | 10 |
| Consumer Goods | 14 |
| Pharma | 10 |
| Financial Services | 14 |
| Energy | 10 |
| Telecom | 3 |
| Metals | 6 |
| Construction | 2 |
| It | 6 |
| Chemicals | 1 |
| Fertilisers & Pesticides | 1 |

**Table 3.** Sector wise distribution of all 89 stocks.

**Data set 2:** We collect the daily data of the constituent stocks of the NIFTY 500 index from the National Stock Exchange (NSE), India, during the 8-year time span from January 2014 to December 2021. The NIFTY 500 index comprises the top 500 companies selected based on full market capitalization from the eligible universe. We removed 149 stocks due to insufficient and missing data, resulting in a total of 351 stocks used for the analysis. Next, we calculate the daily log returns:

$$R_{t+1,k} = ln\frac{\hat{S}_{t+1,k}}{\hat{S}_{t,k}}$$

where $\hat{S}_{t,k}$ is the price of the $k^{th}$ stock at day $t$. Table 4 provides detailed information about sector-wise distribution of the 351 stocks in the NIFTY 500 index considered in our analysis.

| Industry | # of stocks |
|---|---|
| Financial Services | 48 |
| Capital Goods | 36 |
| Healthcare | 33 |
| Chemicals | 29 |
| Fast Moving Consumer Goods | 26 |
| Automobile and Auto Components | 22 |
| Consumer Durables | 21 |
| Information Technology | 18 |
| Oil Gas & Consumable Fuels | 14 |
| Metals & Mining | 14 |
| Services | 12 |
| Consumer Services | 12 |
| Construction Materials | 11 |
| Power | 9 |
| Realty | 9 |
| Telecommunication | 9 |
| Media Entertainment & Publication | 8 |
| Textiles | 7 |
| Construction | 6 |
| Diversified | 5 |
| Forest Materials | 2 |

**Table 4.** Sector wise distribution of all 351 stocks.

**Algorithms for detecting core-periphery structure:** We briefly describe the methods of Rossa et. al.[26] and Rombach et al.[37] for identifying the core-periphery profiles of a given network and assign "coreness" values to each vertex.

In Ref.[26], the intuitive idea of core and periphery is captured by modeling a random walker from vertex $i$ to vertex $j$ as a Markov chain. Here the vertices of the weighted network constitute the set of states of the Markov chain. and the matrix of transition probabilities (from vertex $i$ to $j$ ) $[p_{ij}]$ is defined in terms of the weighted adjacency matrix $[a_{ij}]$ as $p_{ij} = \frac{a_{ij}}{\sum_h a_{ih}}$. In this paper we compute $a_{ij}$ as $\frac{1+\rho_{ij}}{2}$ when $i \neq j$ and $a_{ij} = 0$ otherwise. Here $\rho_{ij}$ is the Pearson correlation coefficient between vertices $i$ and $j$. The goal is to find the largest subset $S$ of vertices such that if the random walker is parked at any vertex belonging to $S$ then at the next instant there is a high probability that it escapes $S$, that is, there is an extremely slim chance of staying inside $S$. This would imply that the connectivity among the vertices in $S$ is possibly non-existent or very weak. So, intuitively $S$ has periphery vertices. Mathematically, the probability that a random walker stays inside $S$ is given by $\varphi_S = \frac{\sum_{i,j\in S}\pi_i p_{ij}}{\sum_{i\in S}\pi_i}$, where $\pi_i > 0$ be the asymptotic probability of being at vertex $i$. This expression, however, greatly simplifies in view of the irreducibility of the Markov chain leading to the condition

$$\pi_j = \sum_{i\in V(G)} \pi_i p_{ij} = \sum_{i\in V(G)} \pi_i \frac{a_{ij}}{\sum_{h\in V(G)} a_{ih}}.$$

For further details, the reader may refer to Ref.[44]. A simple algebraic manipulation yield $\pi_i = \frac{C_D(i)}{\sum_{j\in V(G)} C_D(j)}$. Here $C_D(i) = \sum_h a_{ih}$ stands for the weighted degree of the $i^{th}$ vertex. This simplifies $\varphi_S$ to the expression below and makes it easy to compute: $\varphi_S = \frac{\sum_{i,j\in S} a_{ij}}{\sum_{i\in S, j\in V(G)} a_{ij}}$.

.

Now, the set $S$ is constructed as follows: We start with any vertex of least weighted degree and call the set containing this vertex as $S_1$. Without loss of generality assume that $S_1 = \{1\}$. Here $\phi_1 \coloneqq \phi_{S_1} = 0$. In the next step consider the subsets $S_2^{(j)} \coloneqq S_1 \cup \{j\}$ for all $2 \leq j \leq N$ and compute the minimum of all $\phi_{S_2^{(j)}}$. Suppose $\phi_{S_2^{(k)}}$ is the minimum then define $S_2 \coloneqq S_2^{(k)}$ and $\phi_2 \coloneqq \phi_{S_2^{(k)}} = \phi_{S_2}$ is the "coreness" assigned to the vertex $k$. Note that $S_2$ is a set of 2 vertices with the least persistence probability and $\phi_1 \leq \phi_2$. Next, we repeat the above process to construct $S_3$ from $S_2$ and to compute $\phi_3$ - the “coreness” of the third vertex so that $\phi_1 \leq \phi_2 \leq \phi_3$ and continue this process till all vertices are exhausted. Finally, we have the core-periphery profile $\phi_1 \leq \phi_2 \leq \cdots \leq \phi_N$ of the network.

**Fig. 1** shows the correlation based PMFG network of 89 stocks of the index CNX100 for the entire year 2014. Green vertices are the ones which have high values of coreness and the red vertices have very low values of coreness (that is they are most peripheral). Had there been an ideal core-periphery structure in this network then we would have found a set of red vertices in which no two vertices are connected but that is not the case here as one can spot two adjacent red vertices in the figure.

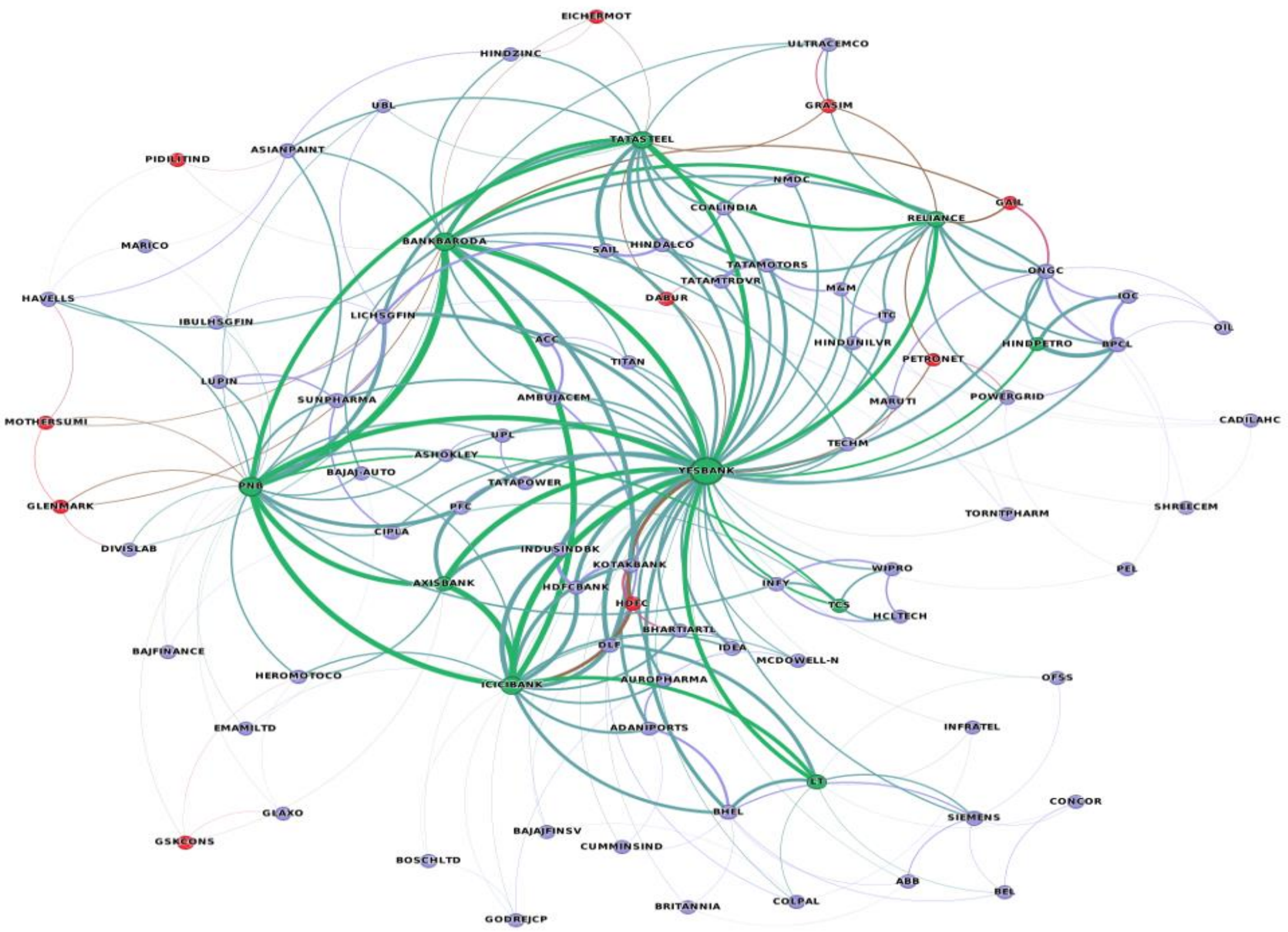

**Figure 7.** A correlation based PMFG network of 89 stocks of the index CNX100 from the National Stock Exchange, India, from January to December 2014. The top 10 peripheral and core stocks based on the method Rossa et al. are colored with red and green circles, respectively. Also, the size of a vertex is proportional to the degree of the vertex, and the width of the edge is proportional to the correlation coefficients.

The other idea of Rombach et al.[37] was to assign a core score (a number lying between 0 and 1) to each vertex through an optimization procedure described below. This method works for weighted and undirected networks and hinges around determining a random shuffling $(c_1, c_2, \dots, c_N)$ of the "coreness" values assigned to $N$ vertices which maximizes the quality function for chosen values of $\alpha$ and $\beta$:

$$Q(\alpha,\beta) = \sum_{i=1}^{N}\sum_{j=1}^{N} a_{ij}\, c_i(\alpha,\beta) c_j(\alpha,\beta).$$

Following Ref.[37], we choose the initial "coreness" values to be

$$c_i^*(\alpha,\beta) = \begin{cases} \dfrac{i(1-\alpha)}{2\lfloor\beta N\rfloor}, & i \in \{1,\dots,\lfloor\beta N\rfloor\}, \\ \dfrac{(i-\lfloor\beta N\rfloor)(1-\alpha)}{2(N-\lfloor\beta N\rfloor)} + \dfrac{1+\alpha}{2}, & i \in \{\lfloor\beta N\rfloor + 1,\dots,N\}. \end{cases}$$

The parameters $\alpha, \beta$ lie between 0 and 1. Let the shuffling of $(c_i^*)$ which maximizes $Q$ be denoted by $(c_i)$. For the sake of convenience, we denote the maximum value of $Q(\alpha,\beta)$ also by $Q(\alpha,\beta)$. We repeat these calculations for numerous pairs $(\alpha,\beta)$ and define the core score (CS) for each vertex $i$ as

$$CS(i) = Z \sum_{\alpha,\beta} c_i(\alpha,\beta) Q(\alpha,\beta), \tag{3}$$

where $Z$ is the normalizing constant chosen so that $\max_{1\le j\le N} CS(j) = 1$.

In this paper we work with 10000 pairs $(\alpha,\beta)$ uniformly sampled from the unit square [0, 1] × [0, 1].

Supplementary Fig. S14 displays the correlation based PMFG network of 89 stocks of the index CNX100 for the entire year 2014. The color of the vertices is according to the core-score given by equation (3).